\documentstyle[12pt,a4wide]{article}
\def\ga{{\ \lower-1.2pt\vbox{\hbox{\rlap{$
          >$}\lower5pt\vbox{\hbox{$\sim$}}}}\ }}
\def\la{{\ \lower-1.2pt\vbox{\hbox{\rlap{$
          <$}\lower5pt\vbox{\hbox{$\sim$}}}}\ }}
\def\he3he4{$^3$He-$^4$He}
\def\eps{\varepsilon}
\def\ua{\uparrow}
\def\da{\downarrow}
\def\su{\uparrow}
\def\sd{\downarrow}

\def\ds{\displaystyle}
\def\g1t{$\tilde{\Gamma}_1$}

\input{psfig.sty}

\begin{document}
\title{\bf
  Superfluid Transition Temperature in a Fermi
Gas with Repulsion. Higher Orders Perturbation Theory Corrections.}
\author{ D.V. Efremov$^{a1}$,
M.S. Mar'enko$^{a2}$,
M.A. Baranov$^{b}$,
M.Yu. Kagan$^{a}$\\
  {\it $^{a}$ P.L. Kapitza Institute for Physical Problems, Moscow
  117334,Russia}\\
  {\it $^{b}$ Russian Research Center "Kurchatov Institute", Moscow 123182,Russia}}

\maketitle
%\addtocounter{footnote}{1}
\footnotetext[1]{E-mail:efremov@kapitza.ras.ru}
\footnotetext[2]{E-mail:maxim@kapitza.ras.ru}

\begin{abstract}
High order perturbation theory corrections to the superfluid transition
temperature in a weakly interacting Fermi gas with repulsive interaction
are calculated. This involves calculating the contributions
of third and fourth order diagrams in the gas parameter  $ap_F$ and taking
into account effects of retardation. The contributions from both second, third
and fourth orders result in the effective attraction in $p$-wave channel.
It  is shown that the critical temperature is mainly determined by second and
third orders terms of perturbation theory. The dependence of the critical
temperature on an
external magnetic field is found.  We discuss possible applications of
the results to the diluted $^3$He-$^4$He  mixtures and trapped
neutral-atom Fermi gases.
 \end{abstract}
PACS:      67.60.-g, 74.20.-z, 74.20.Mn

\section{Introduction}

Unconventional mechanisms of Cooper pairing have
recently started to attract greater attention. This is primarily
related to the discovery of high-temperature
superconducting (HTSC) systems, superconductivity
in organic materials and heavy-fermion compounds,
and also because of the search for superfluidity in
of $^3$He-$^4$He mixtures and in trapped atomic Fermi gases.
However, HTSC and heavy-fermion systems
belong to a class of strongly correlated systems whose
theoretical analysis requires the development of new
methods. At the same time, $^3$He-$^4$He mixtures and
trapped atomic Fermi gases can be described using the
model of a weakly interacting Fermi gas. In this case the
interparticle interaction can be either attractive or
repulsive. In the attractive case conventional singlet Cooper
pairing takes place where the orbital momentum of the
pair is $\ell = 0$, for which the
critical temperature was first calculated by Gor'kov and
Melik-Barkhudarov \cite{Gorkov61}. In systems with repulsive
interaction, the formation of $\ell = 0$ Cooper pairs is
clearly impossible and in order to investigate the existence
of superfluidity, we need to study the possibility of
$\ell \neq 0$ Cooper pairing.
The possible existence of superfluidity in Fermi systems
with repulsion was first indicated by Kohn and
Luttinger in 1965. In \cite{Kohn} they examined the contribution
of collective effects to the scattering amplitude in
a particle-particle channel which lead to effective
quasiparticle interaction at the Fermi surface via polarization
of the Fermionic background. A principal role in the formation
of attractive harmonics in the effective interaction
and consequently the superfluidity is played by the
Kohn singularity in the effective interaction. In the
three-dimensional case, it has the form
\begin{equation} \tilde{\Gamma}_{ \rm eff}^{\rm sing}(q) \sim \left[(2p_F)^2
-q^2 \right] \ln \left| (2p_F)^2 -q^2 \right| + \Gamma_{\rm reg}(q^2).
\end{equation}
In coordinate space the Kohn singularity leads to alternating
sign and magnitude dependent RKKY-type interaction between quasiparticles:
\[ \tilde{\Gamma}_{\rm eff}^{\rm sing}(r) \sim \frac{1}{r^3} \cos(2p_F r +
\varphi).  \]
It should be noted that the above contribution to the effective
interaction decreases over large distances more slowly
that the bare interaction $U_0(r-r')$ and consequently
corresponds to the main contribution to the scattering amplitude
in the limit of large momenta $\ell$ \cite{Kohn}:
\[
\tilde{\Gamma}_{\rm eff}^{(l)} \sim \frac{(-1)}{l^4}.  \]
A simple extrapolation made by the authors of \cite{Kohn}
yields extremely low estimates for superfluid transition
temperatures in the limit $l\rightarrow 2$: $10^{-16}$К and $10^{-11}$К
for $^3$He and the electron subsystem in the metal, respectively.
It was subsequently shown in \cite{Fay68},
\cite{Kagan88} that effective
attraction occurs also for the angular momentum
$\ell=1$ which gives the following expression for the critical
triplet-pairing temperature in the second order of
perturbation theory:
\begin{equation} T_{c1} \sim \tilde{\eps} \exp \left\{
  -\frac{5\pi^2}{4(2\ln 2-1) (ap_F)^2} \right\} \approx \tilde{\eps} \exp
  \left\{ - \frac{13.0}{\lambda^2} \right\} \label{Tc2}.  \end{equation}
where  $\lambda = (2 ap_F)/\pi$ is the gas parameter, $a$ is the
$s$-wave scattering length, $p_F$ is the Fermi momentum, and
$\tilde{\eps}$
is the energy parameter, of the order of the Fermi energy,
which provides a cutoff at high energies.
Substituting experimental values for $^3$He where triplet pairing
takes place, gives good agreement with experiment: $T_{c1} \sim
10^{-3} $K. (Obviously, the bare interaction in real $^3$He is far
more complex than that in the considered model).
The aim of the present paper is to determine the critical
superfluid transition temperature of a weakly non-ideal
Fermi gas with repulsive interparticle interaction
up to the preexponential factor. For this purpose we
calculate the irreducible vertex in the Cooper channel
in the third and fourth orders of perturbation theory
with respect to the gas parameter $\lambda$. We also allow for
renormalization of the singular parts of the Green's functions
(corrections associated with the  $Z$-factor and the effective
mass) in the Bethe-Salpeter equation (\ref{BSE1}) and take
into account retardation effects (the influence of the frequency
and momentum dependencies of the irreducible
vertex).
This article is organized as follows. In Section 2
we derive and analyze an equation for the critical
temperature in a weakly interacting Fermi gas with
repulsion. In Section 3 we calculate the irreducible vertex
in the Cooper channel in the second, third, and
fourth orders of perturbation theory. In Section 4 we
examine the contribution of retardation effects. In Sections
5 and 6 we give the final formula for the critical
temperature and discuss the contribution of the bare
scattering in the $p$-wave channel. In Section 7 we note the
possibility of a strong enhancement of $T_{c1}$ in an external
magnetic field. In Section 8 we discuss possible experimental
applications of the obtained results. In particular, we
discuss the possibility of triplet Cooper pairing in
$^3$He-$^4$He mixtures and in a trapped neutral-atom Fermi gases at ultralow temperatures.

\section{Superfluid transition in a Fermi gas with repulsion}

We  consider a weakly interacting Fermi gas
described by the Hamiltonian
\begin{eqnarray*}
\hat{H} &=& \hat{H}_0+\hat{H}_{\rm int} =
\sum_{\alpha \, {\bf p}}
\left( \eps_{\bf p} - \mu \right)
\hat{a}^{\dagger}_{{\bf p} \alpha}\hat{a}_{{\bf p} \alpha}+ \\ &+&
%\frac{2\pi g\hbar^2}{m V}
\frac{g}{2}
\sum_{\alpha \beta {\bf p p' q}}
\hat{a}^{\dagger}_{{\bf p}\alpha}\hat{a}^{\dagger}_{{\bf p'}\beta}
\hat{a}_{{\bf p'+q}\beta}\hat{a}_{{\bf p-q}\alpha},
\end{eqnarray*}
where the indices $\alpha, \beta=1,2$ label the system components
which we assume to have equal masses $m$ and
concentrations $n_{1,2}=p_F^3/6\pi^2$ , $\mu$ is the chemical potential,
and the constant $g$ characterizes the interparticle
interaction which we shall assume to be point-like (here and
subsequently we put $\hbar =1$). The specific physical
meaning of the
two components depends on the
particular system. For example, for \he3he4 mixtures
 it corresponds to an "upward" and "downward"
projection of the spin, whereas in the case of trapped atomic gas,
 it corresponds to a hyperfine-structure
component (or projection of the nuclear spin).
The considered form of the interparticle interaction assumes
that only $s$-wave scattering takes place in the system, characterized
by the scattering length $a$. (In the leading order
of perturbation theory $a=mg/
4\pi$.) The corresponding
small dimensionless parameter, the gas parameter $\lambda$, is
given by $\lambda = 2 |a|p_F/ \pi$.
We subsequently show how the final result is modified
in the presence of scattering in channels with nonzero
orbital momenta.
It is well known that the appearance of superfluid pairing
is associated with the presence of a pole in the
two-particle vertex function $\Gamma$ in the particle-particle
channel (Cooper channel) for zero total momentum
and frequency \cite{Landau9}. This vertex function $\Gamma$ is a solution
of the Bethe-Salpeter integral equation (Fig. \ref{fig:bethe}):

\begin{figure}[t]
\vspace{0.05\textheight}
\centerline{\begin{picture}(0,0)%
%\centerline{ %
%\hspace{1.5in}
\psfig{file=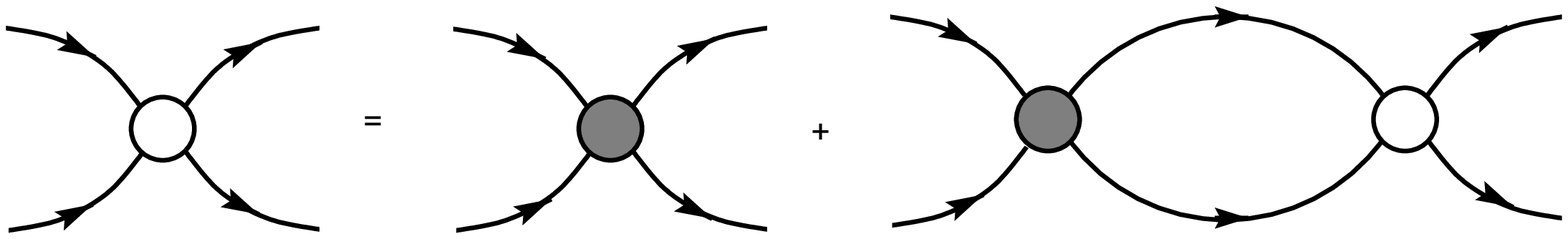,width=6.75in}%}
\end{picture}%
\setlength{\unitlength}{3000sp}%
\begin{picture}(10491,2314)(3118,-4469)
\put(10150,-2761){\makebox(0,0)[b]{$\tilde{\Gamma}$}}
\put(7200,-2761){\makebox(0,0)[b]{$\tilde{\Gamma}$}}
\put(12451,-2761){\makebox(0,0)[b]{$\Gamma$}}
\put(6151,-4411){\makebox(0,0)[lb]{ $-{\bf p}_1$}}
\put(9151,-4411){\makebox(0,0)[lb]{ $-{\bf p}_1$}}
\put(13501,-4411){\makebox(0,0)[lb]{$-{\bf p}_3$}}
\put(13501,-2386){\makebox(0,0)[lb]{${\bf p}_3$}}
\put(11176,-2386){\makebox(0,0)[lb]{${\bf q}$}}
\put(11101,-4411){\makebox(0,0)[lb]{$-{\bf q}$}}
\put(9151,-2386){\makebox(0,0)[lb]{ ${\bf p}_1$}}
\put(6151,-2386){\makebox(0,0)[lb]{ ${\bf p}_1$}}
\put(5101,-2386){\makebox(0,0)[lb]{ ${\bf p}_3$}}
\put(4951,-4411){\makebox(0,0)[lb]{ $-{\bf p}_3$}}
\put(3151,-4411){\makebox(0,0)[lb]{ $-{\bf p}_1$}}
\put(3151,-2386){\makebox(0,0)[lb]{ ${\bf p}_1$}}
\put(3976,-2761){\makebox(0,0)[lb]{ $\Gamma$}}
\put(7951,-4411){\makebox(0,0)[lb]{ $-{\bf p}_3$}}
\put(8101,-2386){\makebox(0,0)[lb]{ ${\bf p}_3$}}
\end{picture}
                       }
                       \vspace{0.1\textheight}
\caption{
   Bethe-Salpeter equation for complete vertex  $\Gamma$.
}
\label{fig:bethe}
\vspace{0.05\textheight}
\end{figure}
\begin{eqnarray} \label{BSE1}
  &&\Gamma(p_1,-p_1;p_3,-p_3)= \tilde{\Gamma}
  (p_1,-p_1;p_3,-p_3) - \\ \nonumber
  && - T \sum_{n=- \infty}^{+\infty} \int
  \tilde{\Gamma} (p_1,-p_1;q,-q) G (\omega_n,{\bf q}) G (-\omega_n,-{ \bf q})
   \Gamma (q,-q;p_3,-p_3) \frac{d^3 q}{(2\pi)^3} ,
\end{eqnarray}
where $\tilde{\Gamma}$ is the irreducible vertex in the Cooper channel
(having no singularities at zero total momentum and
frequency), $G$ is the single-particle Green's function, and
the arguments of the vertex functions denote the corresponding
sets of Matsubara frequencies and momenta:
$ q =(\omega_n, {\bf q}) $, $ p_1 = (\omega_{n1}, {\bf p}_1) $, and so on.
 Note that in formula
(\ref{BSE1}) (and in the following formulas) we do not explicitly indicate
the indices distinguishing the components
of the Fermi gas (for example, $\Gamma$ should be
considered as $\Gamma_{\alpha \beta \gamma \delta}$, and so on). Writing them in
explicit form does not present any difficulties. We also
note that the nonsymmetrized (in terms of the component
indices) irreducible vertex function is used in
equation (\ref{BSE1}).
The vertex functions $\Gamma$ and $\tilde{\Gamma}$ appearing in (\ref{BSE1}) are in
fact functions of the Matsubara frequencies, the absolute values
of the incoming and outgoing momenta, and the angle
between them. For example, we have
$$
\Gamma(p_1, -p_1;p_3, -p_3) = \Gamma(\omega_1, \omega_3,
|{\bf p}_1|, |{\bf p}_3|, \cos(\theta_{{\bf p}_1 {\bf
p}_3})). $$
Thus, expanding G and as a series in terms of Legendre
polynomials
\begin{eqnarray} \label{legandre}
  &&\tilde{\Gamma}(\ldots, \cos(\theta)) = \sum_{l=0}^{+\infty} (2l+1)
  \tilde{\Gamma}_l (\ldots) P_l (\cos \theta) ,\\ \nonumber
  &&{\Gamma}(\ldots, \cos(\theta)) = \sum_{l=0}^{+\infty} (2l+1)
  {\Gamma}_l(\ldots) P_l (\cos \theta)
  \end{eqnarray}
and integrating over angles, we easily obtain from (\ref{BSE1})
the following equation for the singular part $\Gamma^{(s)}_l$ of the $\ell$-th
harmonic of the vertex function:
\begin{eqnarray}\label{BSE2}
 && {\mbox~~~~~~~~~~~~~~~} \Gamma^{(s)}_l(\omega_1,\omega_3,|{\bf p}_1|,|{\bf
 p}_3|)=\\
 &&- T \sum_{n=- \infty}^{+\infty} \int \frac{d^3q}{(2\pi)^3}
  \tilde{\Gamma}_{l}(\omega_1,\omega_n,|{\bf p}_1|,|{\bf q}|)
   G (\omega_n,{\bf q}) G (-\omega_n,-{ \bf q})
   \Gamma^{(s)}_l(\omega_n,\omega_3,|{\bf q}|,|{\bf p}_3|)
   .\nonumber
\end{eqnarray}

As usual, the critical temperature corresponds to the
appearance of a nontrivial solution of this equation
which is related to singular (logarithmic) behavior
of the Cooper loop near the Fermi surface. Thus, in
order to determine the critical temperature in the leading
order in $\lambda$, it is sufficient to set the frequencies to
zero and the absolute values of the momenta to $p_F$ in all the vertex functions
contained in (\ref{BSE2}). We then have
\begin{equation} \label{cooper}
- T \sum_{n}^{} \int \frac{d^3 q}{(2 \pi)^3} G(\omega_n, q)
G(-\omega_n, -q) \rightarrow \frac{m^{*}}{m} Z^2 \nu_F \ln
\frac{\tilde{\eps}}{T_{c1}},
\end{equation}
where $\nu_F=mp_F /2\pi^2$ is the density of states at the
Fermi surface,
$m^*$ is the effective mass, and $Z$ is the residue in
the singular part of the single-particle Green's function. In
equation (\ref{cooper}) $\tilde{\eps}\sim \eps_F$ is the cutoff parameter which
depends on the behavior of $\tilde{\Gamma}$ at high momenta and frequencies.
Now equation (\ref{BSE2}) can be rewritten in the form
\begin{equation}\label{BSE3}
\Gamma^{(s)}_l = \tilde{\Gamma}_l
\cdot Z^2 \frac{m^{*}}{m} \nu_F \ln
\left( \frac{\tilde{\eps}}{ T} \right) \cdot
\Gamma^{(s)}_l,
\end{equation}
where
$\tilde{\Gamma}_l
=\tilde{\Gamma}_l(p_i=p_F,\omega_i=0)$ и $\Gamma_l^{(s)}
= \Gamma_l^{(s)}(p_i=p_F,\omega_i=0)
$,
so that a nontrivial solution is only possible for $\tilde{\Gamma}<0$
and occurs at temperature $T=T_{cl}$ where
\begin{equation} \label{tc_l}
T_{cl} = \tilde{\eps} \exp \left\{ -\frac{1}{\nu_F
|\tilde{\Gamma}_l|} \frac{m}{m^* Z^2} \right\}.
\end{equation}

For the case of a Fermi gas with interparticle attraction
 we have
 $\tilde{\Gamma}_0 \simeq 4\pi a /m <0$,
and the system is unstable with respect to traditional
$s$-wave pairing ($/el=0$). The superfluid transition temperature
 in this case was obtained in  \cite{Gorkov61} to within terms
$O(\lambda^0)$ and is given by
\begin{equation}\label{Tc:gorkov}
  T_{c0}=\frac{1}{\pi} e^C \left( \frac{2}{e} \right)^{7/3} \eps_F \exp
  \left\{ -\frac{\pi}{2|a|p_F}
    \right\} \approx 0.28 \eps_{F} \exp \left\{
    - \frac{1}{\lambda} \right\},
  \end{equation}
where $C = 0.58\ldots$ is the Euler constant. This expression
only differs from the corresponding expression for $T_{c0}$ in BCS
theory in that the Debye frequency $\omega_D$ is replaced by $\eps_F$
in the  preexponential factor. This
replacement means that in our case the
entire Fermi sphere and not only its vicinity of the order
of the Debye frequency, is involved in the pairing. Note
that in order to find the preexponential factor in \cite{Gorkov61} we
need to calculate $\tilde{\Gamma}_0$ to within terms of the second order
of perturbation theory.

\begin{figure}[t]\label{fig:2order}
\vspace{0.05\textheight}
%\centerline{\hspace{7mm} \psfig{file=figures/2order.ps,width=6in}}
\setlength{\unitlength}{4000sp}%
\begin{picture}(6647,3968)(526,-4401)
\put(526,-4301){ \psfig{file=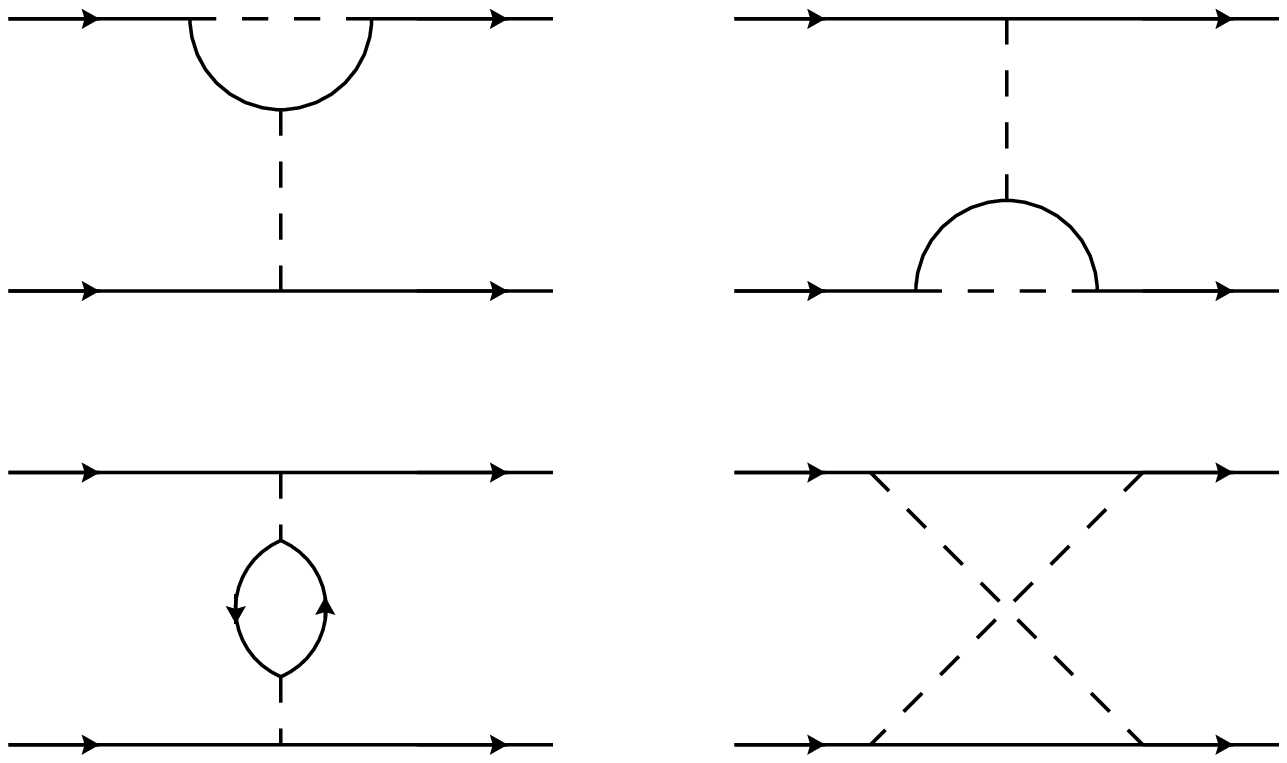,width=6in}}
\put(4556,-2634){${\bf p}_1$}
\put(1206,-2634){${\bf p}_1$}
\put(1102,-2320){$-{\bf p}_1$}
\put(1206,-541){${\bf p}_1$}
\put(4556,-541){${\bf p}_1$}
\put(3247,-2320){$-{\bf p}_3$}
\put(3300,-2687){${\bf p}_3$}
\put(3300,-593){${\bf p}_3$}
\put(6649,-593){${\bf p}_3$}
\put(6545,-2268){$-{\bf p}_3$}
\put(6597,-2687){${\bf p}_3$}
\put(6545,-4361){$-{\bf p}_3$}
\put(4504,-4361){$-{\bf p}_1$}
\put(3247,-4361){$-{\bf p}_3$}
\put(1102,-4361){$-{\bf p}_1$}
\put(4451,-2268){$-{\bf p}_1$}
\put(3876,-1326){$+$}
\put(7173,-1274){$+$}
\put(3876,-3472){$+$}
\put(526,-3524){$+$}
\put(7120,-3472){$ + ({\bf p}_3 \leftrightarrow -{\bf p}_3)$}
\end{picture}

%\vspace{0.1\textheight}
\caption{
  Second-order diagrams in terms of gas parameter for irreducible interaction.}
  \vspace{0.05\textheight}
\end{figure}

For repulsive interaction, $a>0$, equation (\ref{BSE2}) for $\ell=0$
only has a trivial solution and $s$-wave pairing is impossible.
In this case, superfluid pairing will take place in the
channel having the orbital momentum $\ell$ for which $\tilde{\Gamma}_l$ is
negative and has the maximum absolute value. As it is well known
 \cite{Landau3}, the scattering amplitude of slow particles
with the orbital momentum $\ell$ for the short-range
potential has the order of magnitude $ a(a p)^{2l}$, where $p$ is
the particle momentum and $a$ is the $s$-wave scattering length.
Thus, in our particular case the corresponding contribution
 to $\tilde{\Gamma}_l$ with $\ell>0$
from scattering on the bare interparticle potential has a maximum for
$\ell=1$ and has the order $ (ap_F)^3 \sim
\lambda^3$.
At the same time, many-particle effects associated with
polarization processes of the Fermi background in a
Fermi gas have the order $\lambda^2$ and, therefore, dominates for
$l\ge 1$ \cite{Kagan88}. Corresponding diagrams
 for $\tilde{\Gamma}$ in the second order with respect to $\lambda$ are
plotted in Fig. \ref{fig:2order}.
For our particular case of point interaction
the first three diagrams cancel out so that $\tilde{\Gamma}$ is
completely determined by the last exchange diagram
and is given by (we assumed $T=0$)
\[
\tilde{\Gamma}(\omega_1,\omega_3, {\bf p}_1, {\bf p}_3)=
\left( \frac{4 \pi a}{m} \right)^2 \Pi(\omega_1+\omega_3,
{\bf p}_1 + {\bf p}_3),
\]
where
\begin{equation} \label{POL}
\Pi(\Omega,{\bf q})=\int
\frac{d\omega}{2\pi}\int\frac{d^3p}{(2\pi)^3}G(\omega,{\bf
p})G(\Omega+\omega,{\bf p}+{\bf q}) = \int
\frac{d^3p}{(2\pi)^3} \frac{n({\bf p}+{\bf q})-n({\bf
p})}{\Omega+\xi({\bf p}+{\bf q})-\xi({\bf p})}.
\end{equation}
In this expression $n(p)$ is the Fermi particle distribution
function for $T=0$, $\xi(p)=p^2/2m-\mu$.

From formula (\ref{POL}) we can easily obtain an expression
for the irreducible vertex at zero external frequencies
and momenta lying on the Fermi surface. In terms
of the angle $\theta$ between ${\bf p}_1$ and ${\bf
p}_3$ we have
\[
\nu_F\tilde{\Gamma}(0,0,p_1=p_F,p_3=p_F,\cos
\theta)=\frac{\pi \lambda}{2}\left(1-\frac{\lambda}{2}
\left[ 1+\frac{\sqrt{2}(1+\cos \theta)}{4\sqrt{1-\cos
\theta}} \ln \frac{\sqrt{2}+\sqrt{1-\cos
\theta}}{\sqrt{2}-\sqrt{1-\cos \theta}} \right]\right).
\]
As a result of the integration with Legendre polynomials
we obtain
\begin{equation} \label{vert1}
%  && \nu_F \tilde{\Gamma}_0=ap_F\left(1+\frac{2}{3} \frac{ap_F}{\pi} (1+2 \ln2)
%  \right) >0,  \nonumber \\
  \nu_F \tilde{\Gamma}_1=\frac{1}{5} \lambda^2 (1-2
  \ln2)
  <0.
%  \\
%  && \tilde{\Gamma}_2=\frac{4}{105\pi} (ap_F)^2 (11 \ln2-8) \approx 0.1
%  \tilde{\Gamma}_1 <0, \nonumber \\
%  \vspace{0.5em}
%  && \tilde{\Gamma}_3=0.25 \tilde{\Gamma}_2 <0. \nonumber
  \end{equation}
All the other partial components $\tilde{\Gamma}_l$ with $\ell>1$ also correspond
to attraction but are smaller than $\tilde{\Gamma}_1$ and their
absolute values decrease rapidly with increasing $\ell$ (see
\cite{Kagan88}). Thus, we conclude that a weakly interacting Fermi
gas with interparticle repulsion is unstable with respect
to triplet $p$-wave pairing. The corresponding critical temperature
in the leading order with respect to $\lambda$ is
\begin{equation} \label{Tc1}
T_{c1} = \tilde{\eps} \exp \left\{
-\frac{1}{\nu_F |\tilde{\Gamma}_1|}
\right\} =
\tilde{\eps} \exp \left\{
-\frac{5}{(2 \ln2 -1)}\frac{1}{\lambda^2}
\right\}.
\end{equation}

It can be seen from this formula that in order to
determine the preexponential factor $\tilde{\eps}$ in equation (\ref{BSE2})
we need to retain terms up to order $\lambda^4$. (This follows
from the fact that since $\tilde{\Gamma}_1$ begins from terms $\lambda^2$,
to obtain terms of the order $\lambda^0$ in the exponent,
we need to know $\tilde{\Gamma}_1$ up to
terms $\lambda^4$ includingly.)

Note that the contribution of triple collisions can be
neglected within the considered accuracy since it has
the order $\lambda^5$ \cite[\S 6]{Landau9}.

\section{Contribution of higher orders of perturbation theory}

\begin{figure}[t]

\setlength{\unitlength}{3947sp}
\begin{picture}(7725,3948)(301,-4165)
\put(301,-4165){\psfig{file=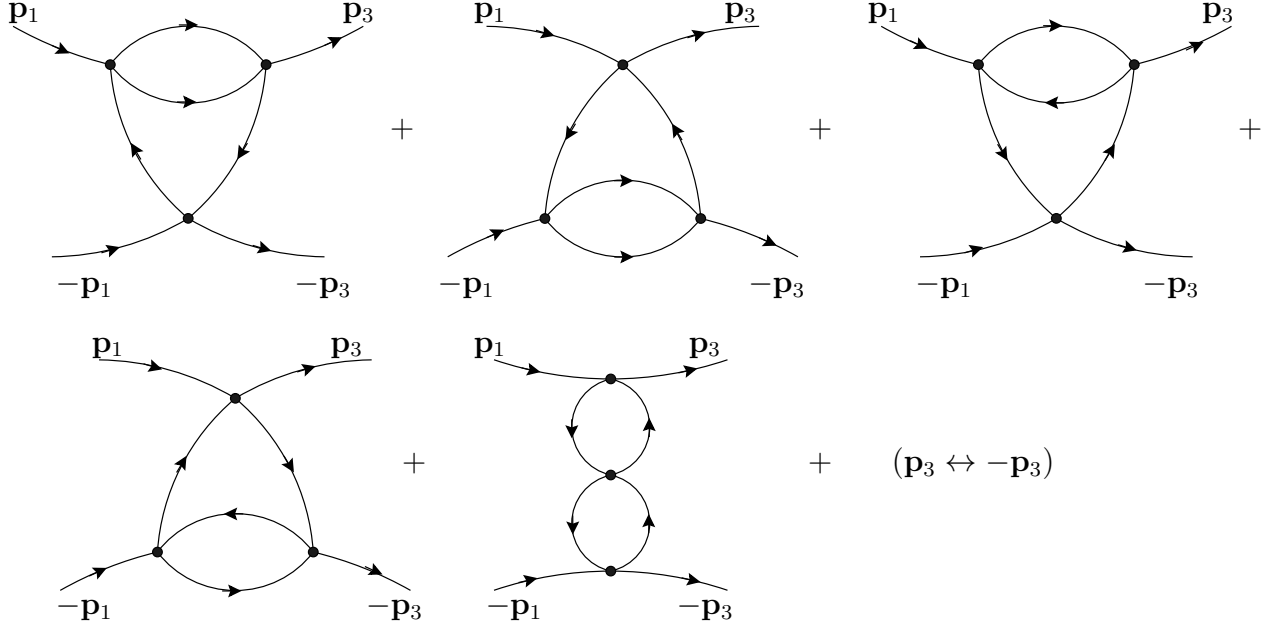,width=6.6in}}
\put(2776,-3211){\makebox(0,0)[lb]{+}}
\put(2701,-1111){\makebox(0,0)[lb]{+}}
\put(5326,-1111){\makebox(0,0)[lb]{+}}
\put(8026,-1111){\makebox(0,0)[lb]{+}}
\put(5326,-3211){\makebox(0,0)[lb]{+}}
\put(5851,-3211){\makebox(0,0)[lb]{(${\rm \bf p}_3
\leftrightarrow -{\rm \bf p}_3$)}}
\put(2326,-2461){\makebox(0,0)[lb]{${\rm \bf p}_3$}}
\put(4576,-2461){\makebox(0,0)[lb]{${\rm \bf p}_3$}}
\put(4951,-2086){\makebox(0,0)[lb]{$-{\rm \bf p}_3$}}
\put(7426,-2086){\makebox(0,0)[lb]{$-{\rm \bf p}_3$}}
\put(3001,-2086){\makebox(0,0)[lb]{$-{\rm \bf p}_1$}}
\put(3301,-4111){\makebox(0,0)[lb]{$-{\rm \bf p}_1$}}
\put(4501,-4111){\makebox(0,0)[lb]{$-{\rm \bf p}_3$}}
\put(2551,-4111){\makebox(0,0)[lb]{$-{\rm \bf p}_3$}}
\put(601,-4111){\makebox(0,0)[lb]{$-{\bf p}_1$}}
\put(826,-2461){\makebox(0,0)[lb]{${\bf p}_1$}}
\put(3226,-2461){\makebox(0,0)[lb]{${\bf p}_1$}}
\put(301,-361){\makebox(0,0)[lb]{${\bf p}_1$}}
\put(2401,-361){\makebox(0,0)[lb]{${\bf p}_3$}}
\put(3226,-361){\makebox(0,0)[lb]{${\bf p}_1$}}
\put(5701,-361){\makebox(0,0)[lb]{${\bf p}_1$}}
\put(7801,-361){\makebox(0,0)[lb]{${\bf p}_3$}}
\put(2101,-2086){\makebox(0,0)[lb]{$-{\bf p}_3$}}
\put(601,-2086){\makebox(0,0)[lb]{$-{\bf p}_1$}}
\put(4801,-361){\makebox(0,0)[lb]{${\bf p}_3$}}
\put(6001,-2086){\makebox(0,0)[lb]{$-{\bf p}_1$}}
\end{picture}
\caption{ Skeleton diagrams of the third order of perturbation
theory for the irreducible vertex $\tilde{\Gamma}$}
\label{fig:3order}
%\vspace{0.1\textheight}
\end{figure}

\begin{figure}[t]
%\vspace{0.1in}

\centerline{\hspace{7mm}
\psfig{file=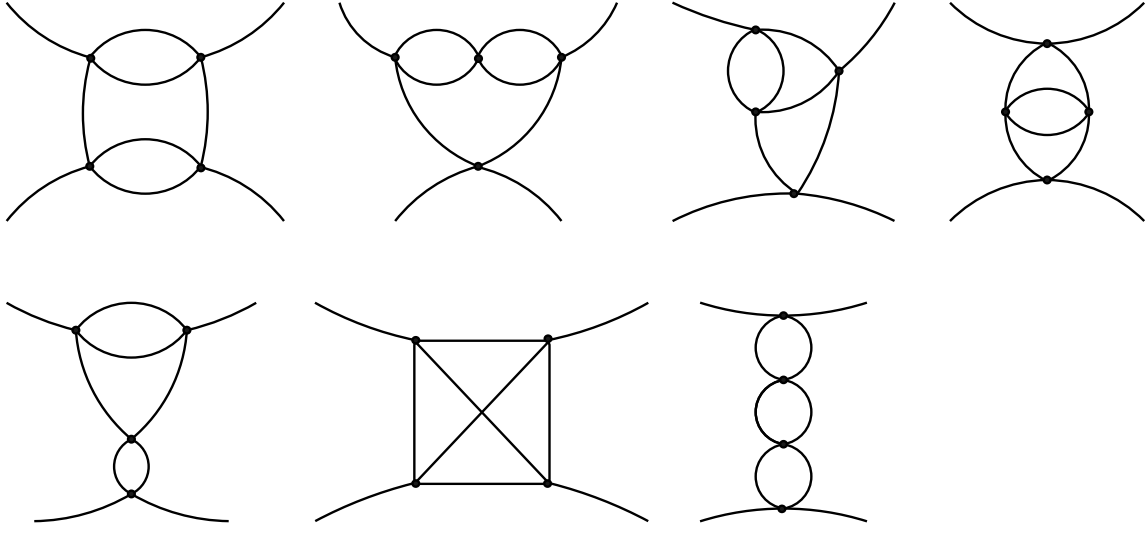,width=6in}}
\caption{Skeleton "nonoriented" diagrams of the fourth order of
perturbation theory for the irreducible vertex $\tilde{\Gamma}$.}
\label{fig:4order}
\vspace{0.1in}
\end{figure}

The irreducible vertex $\tilde{\Gamma}$ in the third and fourth
orders of perturbation theory is given by the diagrams
shown in Figs. \ref{fig:3order} and \ref{fig:4order}, respectively. The points on
these diagrams correspond to antisymmetrized two-particle
interaction. In expanded notation when the
interaction is represented as a dashed line (as in Fig. \ref{fig:2order})
these corresponds to two different ways of connecting
the incoming and outgoing lines.

\begin{figure}[t]
\vspace{0.3in}
% \centerline{\hspace{7mm}
%\psfig{file=figures/4or_ex.ps,width=6in}
%}
\setlength{\unitlength}{3947sp}
\centerline{
\begin{picture}(6181,3525)(773,-3301)
\put(773,-3301){\psfig{file=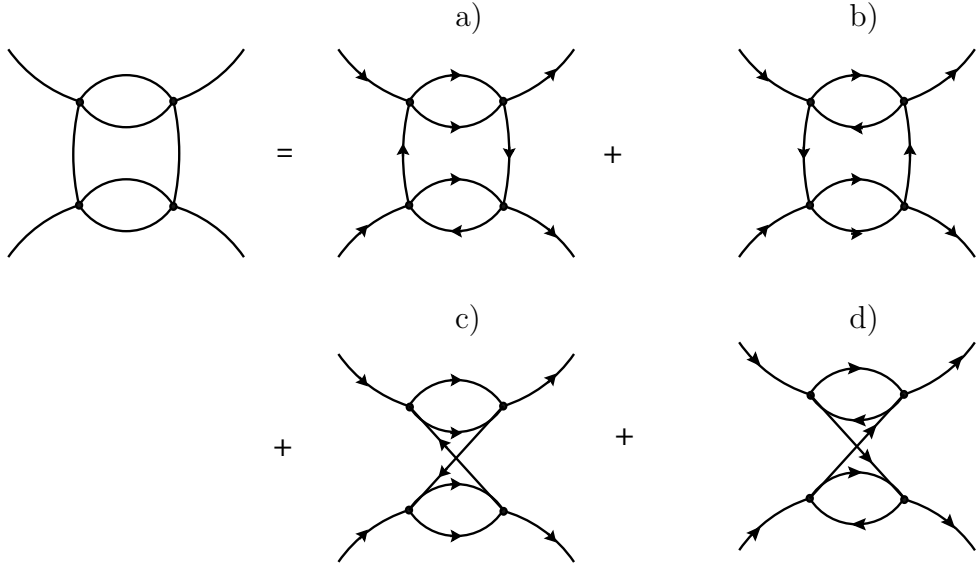,width=0.8\textwidth}}
\put(3600, 89){a)}
\put(6076, 89){b)}
\put(6076,-1810){d)}
\put(3600,-1810){c)}
\end{picture}}
 \caption{Example of decoding "nonoriented" diagrams (first
diagram in Fig. 4).}
 \label{fig:4order_ex} \vspace{0.3in}
\end{figure}

Figures \ref{fig:3order} and \ref{fig:4order} only give skeleton diagrams
(without the self-energy insertions) and Fig. \ref{fig:4order}
only gives "nonoriented" diagrams. The corresponding
Feynman diagrams are obtained by arranging the
arrows on the lines (taking into account the particle number conservation
at the vertexes) and also the incoming and
outgoing momenta. Figure \ref{fig:4order_ex} shows an example of such
an arrangement.

\begin{figure}[t]
\begin{picture}(0,0)
\psfig{file=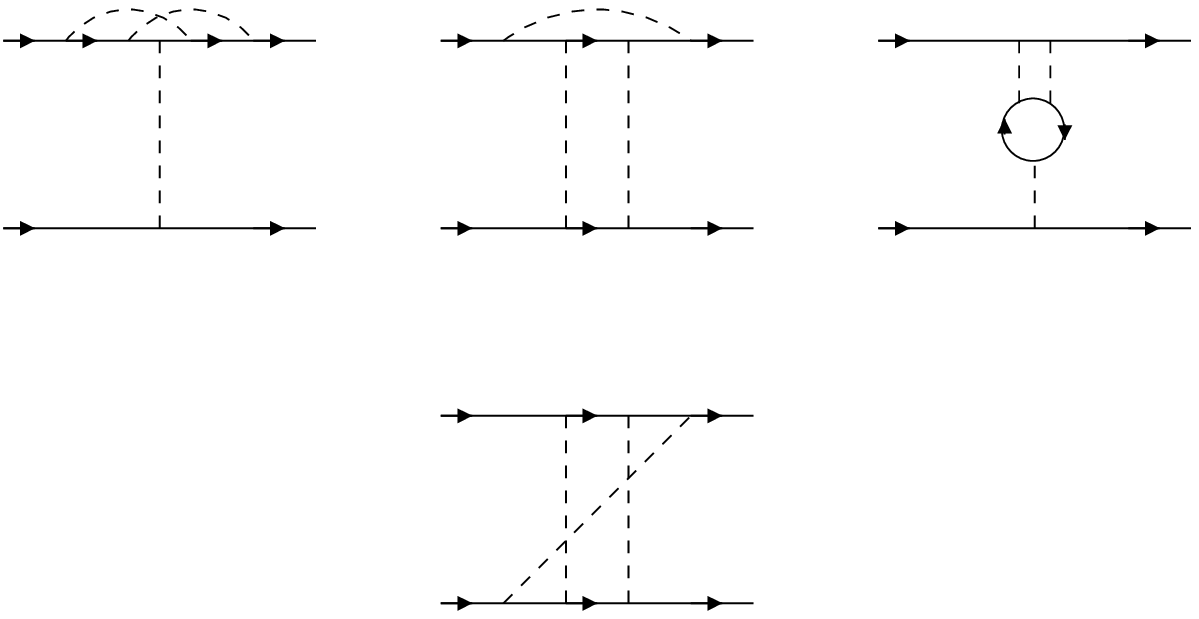,width=5.1 in}
\end{picture}
\setlength{\unitlength}{3947sp}%
\begin{picture}(6012,3444)(889,-3811)
\put(3001,-2311){\makebox(0,0)[lb]{$ {\bf p}_1$}}
\put(4201,-2311){\makebox(0,0)[lb]{$ {\bf p}_3$}}
\put(3001,-3661){\makebox(0,0)[lb]{$ {\bf p}_2$}}
\put(4201,-3661){\makebox(0,0)[lb]{$ {\bf p}_4$}}
\put(2701,-1111){\makebox(0,0)[lb]{$+$}}
\put(4801,-1111){\makebox(0,0)[lb]{$+$}}
\put(6901,-1111){\makebox(0,0)[lb]{$+$}}
\put(3001,-511){\makebox(0,0)[lb]{$ {\bf p}_1$}}
\put(5101,-511){\makebox(0,0)[lb]{$ {\bf p}_1$}}
\put(4201,-511){\makebox(0,0)[lb]{$ {\bf p}_3$}}
\put(6301,-511){\makebox(0,0)[lb]{$ {\bf p}_3$}}
\put(3001,-1861){\makebox(0,0)[lb]{$ {\bf p}_2$}}
\put(5101,-1861){\makebox(0,0)[lb]{$ {\bf p}_2$}}
\put(4201,-1861){\makebox(0,0)[lb]{$ {\bf p}_4$}}
\put(6301,-1861){\makebox(0,0)[lb]{$ {\bf p}_4$}}
\put(901,-1861){\makebox(0,0)[lb]{$ {\bf p}_2$}}
\put(2101,-1861){\makebox(0,0)[lb]{$ {\bf p}_4$}}
\put(2101,-511){\makebox(0,0)[lb]{$ {\bf p}_3$}}
\put(901,-511){\makebox(0,0)[lb]{$ {\bf p}_1$}}
\put(2776,-2911){\makebox(0,0)[lb]{$+$}}
\put(1651,-2011){\makebox(0,0)[lb]{а)}}
\put(5851,-2011){\makebox(0,0)[lb]{c)}}
\put(3751,-3811){\makebox(0,0)[lb]{d)}}
\put(3751,-2011){\makebox(0,0)[lb]{b)}}
\end{picture}
\caption{
First third-order diagram from Fig. 3 showing diagram
corresponding to the substitution ${\bf p}_3
\longrightarrow -{\bf p}_3$, in
expanded representation.}
\label{fig:3or_ex}
 \vspace{0.3in}
\end{figure}

Direct calculations of these diagrams using standard
rules of the diagram technique yield diverging expressions
as a result of the integration over large momenta in
subdiagrams containing Cooper loops (loops formed
from two lines in the same direction). For example,
we consider the first third-order diagram in Fig. \ref{fig:3order}
together with the corresponding diagram in which $p_3$ is replaced with $-p_3$. In
expanded form they correspond to the sum of the diagrams
in Fig. \ref{fig:3or_ex} where the dashed line is the
interparticle interaction. It is easily to check that for
short-ranged potential the first three diagrams cancel out
leaving only the fourth diagram which contains a
subdiagram diverging at large momenta, corresponding
to a Cooper loop between two parallel dashed lines.
However,  this subdiagram is the
first correction of a ladder series to the one of the dashed
lines on the fourth diagram in Fig. \ref{fig:2order} which gives a contribution
to in the second order with respect to $\lambda$.
In the same way, the second diagram in Fig. \ref{fig:3order}  together with the
 corresponding diagram in which $p_3$ is replaced with $-p_3$
  in the sum is the correction to the second
dashed line on the same diagram in
Fig. \ref{fig:2order}.

\begin{figure}[t]

\setlength{\unitlength}{3947sp}%
\centerline{
\begin{picture}(2216,1923)(301,-2140)
\put(301,-2190){\psfig{file=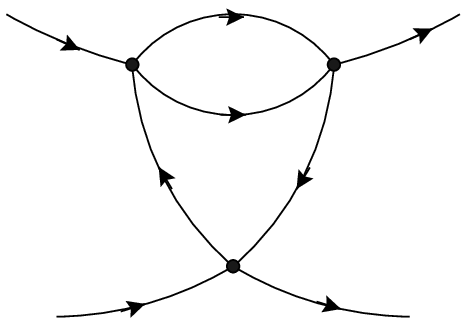,width=2in} }
\put(301,-361){\makebox(0,0)[lb]{${\bf p}_1$}}
\put(2401,-361){\makebox(0,0)[lb]{${\bf p}_3$}}
\put(601,-2086){\makebox(0,0)[lb]{$-{\bf p}_1$}}
\put(1876,-2086){\makebox(0,0)[lb]{$-{\bf p}_3$}}
\put(576,-1261){\makebox(0,0)[lb]{\scriptsize $\Omega,{\bf
\scriptsize{q}}-{\bf \scriptsize{s}}$}}
\put(1876,-1261){\makebox(0,0)[lb]{\scriptsize $\Omega,{\bf
\scriptsize{q}}+{\bf \scriptsize{s}}$}}
\put(1101,-1111){\makebox(0,0)[lb]{\scriptsize
$\frac{\Omega}{2}-\omega,-{\bf \scriptsize{p}}+{\bf
\scriptsize{w}}$}}
\put(1101,-361){\makebox(0,0)[lb]{\scriptsize
$\frac{\Omega}{2}+\omega,{\bf \scriptsize{p}}+{\bf
\scriptsize{w}}$}}
\end{picture}}
\caption{
  Diagram of the third order of perturbation theory
containing a Cooper loop as subdiagram.}
  \label{fig:3glass}
\vspace{0.1in}
\end{figure}

These corrections only differ from the first term of
the Born series for the scattering amplitude in that they
contain the single-particle Green's functions in the
medium $G$ and not in vacuum $G^{(0)}$. However, at large
momenta the difference between $G$ and $G^{(0)}$ disappears
so that the divergence in the diagram in Fig. \ref{fig:3glass} can be
eliminated by changing from the bare interaction $g$ to
the scattering length $a$ (renormalization procedure).
This scattering length $a$ is determined by the scattering amplitude
of two particles in vacuum in the limit where the energies
of the colliding particles tend to zero and can be obtained from the equation
\begin{eqnarray} \label{g->a:int}
\frac{4\pi a}{m}&=& g + \int \frac{d\omega}{2\pi} \int
\frac{d^3p}{(2\pi)^3} g  G^{(0)}(\omega,p)
G^{(0)}(-\omega,-p) \frac{4\pi a}{m}=\\ \nonumber &=&
g+\int \frac{d^3p}{(2\pi)^3} g \frac{1}{2 \eps(p)+i 0}
\frac{4\pi a}{m}.
\end{eqnarray}

After integrating in diagram Fig. \ref{fig:3or_ex}d over the intermediate
frequency of the Cooper loop $\omega$, we obtain the
expression
\begin{eqnarray} \label{plan:renorm}
\int \frac{d^4q}{(2\pi)^4} \int \frac{d^3
p}{(2\pi)^3}\left( \frac{1-\theta(\xi_1)-\theta(\xi_2)}
  {\Omega-(\xi_1+\xi_2)+i\delta(\hbox{sign}\xi_1+\hbox{sign}\xi_2)}
  \right)\times && \nonumber\\
   \frac{1}{(\Omega-\xi_3+i\delta \hbox{sign}\xi_3)(\Omega-\xi_4+i\delta
  \hbox{sign}\xi_4)},&&
\end{eqnarray}
where
$\xi_1=\xi({\bf p}+\frac{{\bf q}+{\bf w}}{2})$,
$\xi_2=\xi(-{\bf p}+\frac{{\bf q}+{\bf w}}{2})$,
$\xi_3=\xi({\bf q}-{\bf s})$, $\xi_4=\xi({\bf q}+{\bf s})$,
${\bf p}_1={\bf s}+{\bf w}$, ${\bf p}_3={\bf s}-{\bf w}$.
The integral over the internal momentum $\bf{p}$ of
the expression in brackets diverges at large momentum. As we have already
noted, this divergence is of the same type as the Born correction to the
scattering amplitude. Therefore, the divergence in the considered third
order diagram can be eliminated by replacing the bare interaction constant
$g$ with the scattering length $a$ in the second order diagram in
Fig. \ref{fig:2order}d.

To within the required accuracy the relationship
between $g$ and $a$ can be obtained from (\ref{g->a:int}) and has the form
 \begin{eqnarray} \label{g->a:2}
  g&=&\frac{4 \pi}{m} a + \left(
 \frac{4 \pi}{m} a\right)^2
  \int
  \frac{d^3 p d \omega }{(2\pi)^4} G^{(0)}(\omega,{\bf p})G^{(0)}(-\omega,-{\bf p})
\\
  &=&\frac{4 \pi}{m} a + \left(
 \frac{4 \pi}{m} a\right)^2
  \int
  \frac{d^3 p}{(2\pi)^3}  \frac{1}{2 \eps (p) +i\delta} ,
  \nonumber
  \end{eqnarray}
where $\ds \eps(p) = \frac{p^2}{2m}$. The renormalization procedure is
shown schematically as follows:
\begin{equation}
\nu_F \{g^2 \Pi +2g^3 (GG) \tilde{\Pi} \}
\rightarrow \lambda^2 \Pi + 2\lambda^3
[(GG)-(G^{(0)}G^{(0)})] \tilde{\Pi} ,
\end{equation}
where $(GG)$ corresponds to the first cofactor in formula
(\ref{plan:renorm}), $\tilde{\Pi}$ corresponds to the second cofactor,
$$(G^{(0)}G^{(0)})= \int \frac{d \omega}{2 \pi} G^{(0)}(\omega,-{\bf p})
G^{(0)}(\omega,-{\bf p}) = \frac{1}{2 \eps(p) + i\delta},
 $$
and the factor 2 in second term in (16) results from the the
contribution of the second diagram in Fig. \ref{fig:3order}. As we can
easily see, the expression for $\tilde{\Pi}$, being integrated over
frequency, gives the polarization operator $\Pi$
which appears in the first term in formula (16). The last
term in formula (16) can be explicitly written as
\begin{eqnarray}\label{C-C0}
  \Gamma^{(3a)}&=& -2i\left(\frac{2\pi^2 \lambda}{mp_F} \right)^3
  \int \frac{d^4 q}{(2\pi)^4}
  \int \frac{d^3 p}{(2\pi)^3}\frac{1}{(\Omega-\xi_3+i\delta \hbox{sign}\xi_3)(\Omega-\xi_4+i\delta
  \hbox{sign}\xi_4)}\\ \nonumber
  &\times&\left( \frac{1-\theta(\xi_1)-\theta(\xi_2)}
  {\Omega-(\xi_1+\xi_2)+i\delta(\hbox{sign}\xi_1+\hbox{sign}\xi_2)} -\frac{1}{2\eps(p)+i\delta}
  \right) .
 \end{eqnarray}

\begin{figure}[t]
\vspace{0.05\textheight}
%\centerline{\begin{picture}(0,0)%
%\centerline{ %
%\hspace{1.5in}
\setlength{\unitlength}{4144sp}
\begin{picture}(4074,3454)(439,-3313)
\put(439,-3313){\psfig{file=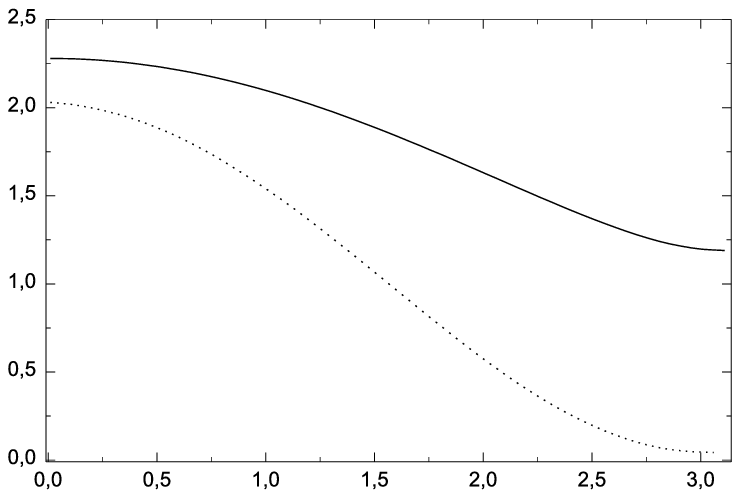,width=0.75\textwidth}}
\put(5096,-3066){\makebox(0,0)[lb]{$\theta_{1-3}$, {\sl rad}}}
%\put(721,100){\makebox(0,0)[lb]{$$}}
\end{picture}
%\end{picture}%
%}
\caption{
Dependence of the irreducible vertex in the second and
third orders on the angle between the incoming and outgoing
momenta $\theta_{1-3}$: $\Gamma^{(2)}/\lambda^2$ -- solid curve,
$\Gamma^{(3)}/\lambda^3$ -- dashed curve.
} \vspace{0.01\textheight}\label{fig:g2g3p}
\end{figure}

This expression contains no divergences and can be
integrated numerically. It can be checked that all other
 third-order diagrams contain no divergences and,
as a result of numerical calculations, we obtain the final
result for the third-order contribution to the $p$-wave harmonic
of the irreducible vertex:
\begin{equation}\label{corr_3order}
  \nu_F \tilde{\Gamma}^{(3)}_1 = -0.33 \lambda^3.
\end{equation}
It should be noted that formula (\ref{corr_3order}) contains no contribution
from Hartree-Fock self-energy components in
the second-order diagrams since this contribution corresponds
to renormalization of the chemical potential.
We also note that the appearance of a large numerical
coefficient 0.33 (compared with the coefficient of 0.077
for the second-order contribution) is associated with
the stronger angular dependence of the third-order diagrams
(see Fig. \ref{fig:g2g3p}). This dependence is mainly determined
by the first two diagrams in Fig. \ref{fig:3order} and can be
attributed to the existence of subdiagrams with Cooper
loops.

All divergences in the fourth-order diagrams can be eliminated
 in exactly the same way. For this purpose, in
the third-order diagrams we need to express $g$ in terms of
$a$ in accordance with formula (\ref{g->a:2}) and in the second-order
diagrams, express $g$ in
terms of $a$ allowing for the term $\sim a^3$, which can easily
be obtained from equation (\ref{g->a:int}). (This term is required
to eliminate the divergences in the second diagram in
Fig. \ref{fig:4order}.) As a result, the contribution of the fourth-order
diagrams in Fig. \ref{fig:4order} is given by
\begin{equation}\label{corr_4order}
  \nu_F \tilde{\Gamma}^{(4)}_1 = -0.39 \lambda^4.
\end{equation}

In order to calculate $\tilde{\Gamma}_1$ to within $\lambda^4$ we also need to
allow for the contribution of the self-energy insertions
of the second order in $\lambda$ in the second-order
diagrams $\tilde{\Gamma}_1^{(2)}$, see Fig. 2. These contributions can no
longer be reduced to renormalization of the chemical
potential. They also result in the appearance of a non-trivial
$Z$-factor and the effective mass $m^{*}$ \cite{AGD}:
\begin{equation}\label{z-factor}
Z=1-\lambda^2 {\ln 2}  ; \; \; \;
\frac{m^*}{m}=1+\frac{2}{15} (7 \ln 2 -1) \lambda^2
\end{equation}
in the singular part of the single-particle Green's function
which now also contains a regular part proportional to
$\lambda^2$. By means of direct numerical calculations of the
corresponding diagrams we can establish that the contribution
of the latter to $\tilde{\Gamma}_1$ is negligible. Thus, we
finally obtain the following expression with the
required accuracy in terms of $\lambda$ for the irreducible vertex
in a Cooper channel with orbital momentum $\ell=1$:
\begin{eqnarray}\label{gamma1_full}
  \nu_F Z^2 \frac{m^{*}}{m} \tilde{\Gamma}_1 &=& \nu_F Z^2 \frac{m^{*}}{m}\left( Z^2
   \frac{m^{*}}{m}
  \tilde{\Gamma}_1^{(2)}+ \tilde{\Gamma}_1^{(3)}+
  \tilde{\Gamma}_1^{(4)}\right) \nonumber \\
  &=&  -0.077\lambda^2-0.33\lambda^3 - 0.26\lambda^4.
\end{eqnarray}

\section{The retardation effects}

In order to determine the critical temperature in Section
2, in equation (\ref{BSE2}) we replaced the irreducible vertex $\tilde{\Gamma}_1$,
which is a function of the incoming and outgoing
frequencies and absolute values of momenta $\tilde{\Gamma}_1(\omega_{i},{\bf
p}_i)$, by
its value at zero frequencies and momenta lying on the
Fermi surface $\tilde{\Gamma}_1(\omega_{i}=0,{\bf
p}_i=p_F)$. In this section we
shall show that allowance for the difference between
$\tilde{\Gamma}_1(\omega_{i},{\bf
p}_i)$ and $\tilde{\Gamma}_1(\omega_{i}=0,{\bf
p}_i=p_F)$ (retardation effects)
introduces a correction of the order of $\lambda^4$ to the vertex
$\tilde{\Gamma}_1$. In other words, these effects influence the numerical
coefficient in the preexponential factor.

Retardation effects are most conveniently taken into
account using a method proposed in \cite[гл. 2]{Ginzburg}.
Omitting the appropriate procedures, which are a trivial
generalization of the derivation of \cite{Ginzburg} to the case of
$p$-wave pairing, we arrive at the following integral equation:
\begin{equation}\label{G5}
  \Phi_1(\xi) =
  -\int\limits_{-\varepsilon_F}^{\infty}d\xi'  \frac{\mbox{th}\,\xi'/2T}{2 \xi'}
  R_1(\xi,\xi') \Phi_{1}(\xi'),
\end{equation}
for which the condition for existence of a nontrivial solution
determines the critical temperature $T_{c1}$. The unknown
function $\Phi_1$ in (\ref{G5}) can be related to the spectral density of
the anomalous Green's function (or more accurately to its first
harmonic in the expansion in terms of Legendre polynomials),
and the kernel $R_1(\xi,\xi')$ is given by
\begin{equation}\label{G4}
  R_1(\xi,\xi') =
  \frac{m}{4\pi^{2}p^2(\xi)}
  \int\limits^{p(\xi)+p(\xi')}_{|p(\xi)-p(\xi')|}\,q d q
\int\limits^{\infty}_{0} \frac{d E
\,\sigma(E,m)}{E+|\xi|+|\xi'|}
\frac{p^2(\xi)+k^2(\xi')-q^2}{2k(\xi')},
\end{equation}
where $\sigma(E,q)$ is related to $\tilde{\Gamma}(\omega=\omega_1-\omega_3,{\bf q}=
{\bf p}_1-{\bf p}_3)$
by
\begin{equation} \label{G2}
\tilde{\Gamma}(i \omega_n, {\bf p}) = \int\limits_{0}^{\infty}
\frac{dE^2\, \sigma (E,{\bf p})}{E^2+\omega_{n}^{2}},
\end{equation}
and the factor $(p^2+k^2-q^2)/2 p k$ is precisely the cosine
of the angle between the incoming and outgoing
momenta which picks up the first harmonic in the
expansion (\ref{legandre}) in terms of Legendre polynomials.

Dividing the region of integration over $\xi'$ in equation
(\ref{G5}) into three parts: $|\xi'| \leq z \eps_F$,
$-\eps_F<\xi'<-z\eps_F$, and $\xi'>z\eps_F$,
where $z$ is an arbitrary number satisfying the condition
$T_{c1} \ll z
\eps_F \ll \eps_F$, and integrating by parts (where the
dependence on $\xi'$ in $R(\xi,\xi')$ and $\Phi_1(\xi')$ can be neglected
in the first region and the hyperbolic tangent in the second
and third regions can be replaced by $\mp 1$, respectively),
equation (\ref{G5}) can be reduced to the form
\begin{equation}\label{G6b}
  \Phi_1(\xi) =
  -\ln\left( \frac{2\gamma\varepsilon_F}{\pi T_{c1}} \right)
  \Phi_1(0)R_1(\xi,0)+\frac{1}{2}
  \int\limits_{-\varepsilon_F}^{\infty} d\,\xi'
  \ln \left(\frac{|\xi'|}{\varepsilon_F} \right)
  \frac{d}{d\,\xi'}
  \left(R_1(\xi,\xi')\Phi_1(\xi')\right).
\end{equation}
(As was to be expected, the arbitrary constant $z$ was
dropped from this equation.) We introduce the new
variable
\begin{equation}\label{G6c}
  \chi(\xi)=\frac{\Phi_1(\xi)}{\Phi_1(0)\ln \ds \frac{\pi T_{c1}}{2\gamma
  \varepsilon_F}},
\end{equation}
which allows us to write the expression for the critical
temperature in the form
\begin{equation}\label{G5d}
  T_{c1}= \frac{2 e^{C} }{\pi} \eps_F
  \exp \left( -\frac{1}{\chi (0)} \right),
\end{equation}
where the function $\chi(\xi)$ satisfies
\begin{equation}\label{G6}
  \chi(\xi)=R_1(\xi,0)+\frac{1}{2}\int\limits^{\infty}_{-\varepsilon_F}
  d\,\xi' \ln\left( \frac{|\xi'|}{\varepsilon_F}
\right)  \frac{d }{d \,\xi'}
\left( R_1(\xi,\xi') \chi(\xi') \right)
\end{equation}

Since the kernel $R_1$ contains the small parameter
($R \sim
\lambda^2$), equation (\ref{G5}) can be solved by an iterative
method. In the zeroth approximation we set:
$$
\chi^{(0)}(\xi)=R_1(\xi,0).
$$
The first correction $\chi^{(1)}$  is given by the integral on the
right-hand side of (\ref{G6}) with $\chi=\chi^{(0)}$:
\begin{equation}\label{G6a}
  \chi^{(1)} = \frac{1}{2}\int \limits_{-\eps_F}^{\infty}
  d\,\xi' \ln\left( \frac{|\xi'|}{\varepsilon_F}
\right)  \frac{d }{d \,\xi'}
\left( R_1(\xi,\xi') R_1(\xi',0) \right)
\end{equation}
and, as can easily be seen, begins with terms of the
order $\lambda^4$. The leading term with respect to $\lambda$ in $\chi_1$ is
obtained if only the leading ($\sim \lambda^2$) terms are retained in
the kernel $R_1$ in formula (\ref{G6a}). In this case, the spectral
function $\sigma(E,{\bf q})$ is the same as the imaginary part of the
operator
\begin{eqnarray*}
  \sigma(E,{\bf q})& =& -\frac{1}{\pi} {\rm Im} \Pi(E,{\bf
  q})= \nonumber \\
  &=& -\frac{1}{\pi} \left\{
  \begin{array}{lcl}
     \displaystyle
     -\frac{m p_F}{4\pi\tilde{q}}
     \left[ 1-\left(\frac{\tilde{E}}{\tilde{q}} -\frac{\tilde{q}}{2}
           \right)^2 \right] &\mbox{~~at~~}&  \displaystyle
           \left|\frac{\tilde{q}^2}{2}-\tilde{q} \right| \le
           \tilde{E}\le \frac{\tilde{q}^2}{2}+\tilde{q}
           \\&&\\
           \displaystyle
 -\frac{m p_F}{4\pi\tilde{q}}2\tilde{E}
 &\mbox{~~at~~}&  \displaystyle
           0 \le
           \tilde{E}\le \tilde{q}
           -\frac{\tilde{q}^2}{2},
  \end{array} \right.
\end{eqnarray*}
where $\tilde{q} = q/p_F$, а $\tilde{E} = E m/p_F^{2}$.
Direct calculations using formula (\ref{G6a}) give
$\chi^{(1)} \approx 0.004 \lambda^4$
which is equivalent to adding $\Delta\tilde{\Gamma}_1$ in formula (\ref{tc_l}),
\begin{equation}\label{retardation}
  \nu_F \Delta \tilde{\Gamma}_1 \approx 0.004 \lambda^4.
\end{equation}
Note that a similar estimate of the contributions of
retardation effects was made in \cite{Alexandrov92} where the authors
used the step function approximation for the frequency
dependence of the polarization operator.
\begin{figure}[t]
\setlength{\unitlength}{4144sp}
%\centerline{
\begin{picture}(4074,3454)(439,-3313)
\put(439,-3313){\psfig{file=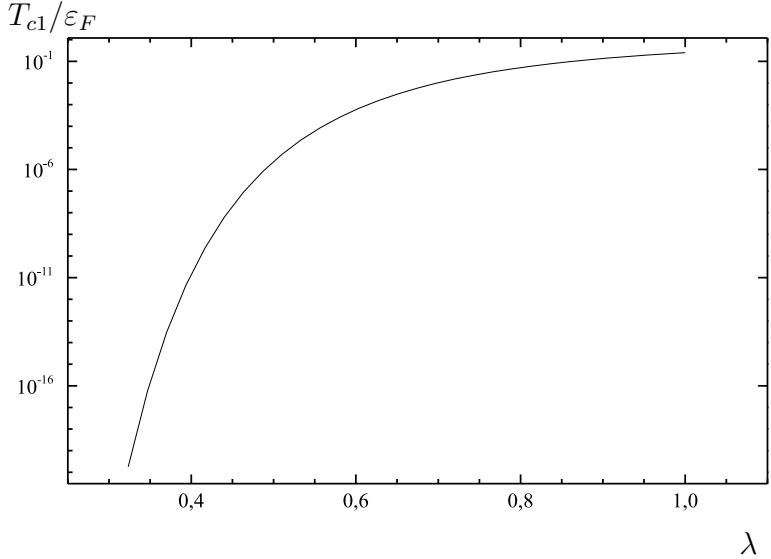,width=0.75\textwidth}}
\put(5096,-3166){\makebox(0,0)[lb]{$\lambda$}}
\put(721,0){\makebox(0,0)[lb]{$T_{c1}/\varepsilon_F$}}
\end{picture}
%}
%\psfig{file=figures/tclambda.ps,width=0.8\textwidth}%}
\caption{
Dependence of $T_{c1}/ \eps_F$ on the gas parameter $\lambda$.}
\label{fig:tc_lambda}
\end{figure}

\section{Critical temperature $T_{c1}$}

Collecting above results together [formulas (\ref{gamma1_full}) and
(\ref{retardation})], we obtain the following expression for the critical
temperature which is determined numerically to
within two decimal places:
\begin{eqnarray}\label{Tc_all}
  T_{c1} &=& \frac{2}{\pi} e^{C}\eps_F \exp \left\{ -
  (0.077
  \lambda^2+0.33\lambda^3+
  0.26 \lambda^4)^{-1}
  \right\} \longrightarrow  \nonumber \\
  \label{Tc_all:expand}
  &&\stackrel{\lambda\rightarrow 0}{\longrightarrow}
  \frac{2}{\pi} e^{C}\eps_F \exp \left\{
    -\frac{13.0}{\lambda^2}+\frac{42.0}{\lambda} - 190
  \right\},
\end{eqnarray}
where the omitted terms have the order $\lambda$. This last formula
assumes $\lambda
<
0.23$ since for $\lambda = 0.23$ the second-
and third-order terms with respect to $\lambda$ in the exponential
function in (\ref{Tc_all}) are the same. For $0.23 \leq \lambda \leq 1$ the
fourth-order term with respect to $\lambda$ in (\ref{Tc_all}) is
smaller than the first two so that (\ref{Tc_all}) can accurately be
rewritten in the form
\begin{equation}\label{Tc_all:expand2}
  T_{c1} \approx \frac{2}{\pi} e^{C}\eps_F \exp \left\{
    -\frac{13.0}{\lambda^2(1+4.3\lambda)}+\frac{42.0}{(1+4.3\lambda)^2}
    \right\}
\end{equation}
This formula may be considered as an extrapolation of
the expression for the critical temperature from $\lambda \ll 1$
[formula (\ref{Tc_all:expand})] to the region $\lambda < 1$
 [formula (\ref{Tc_all:expand2})].
 The dependence $T_{c1}(\lambda)$ is shown in Fig. \ref{fig:tc_lambda}.

\section{Influence of bare $p$-wave scattering}

So far we have only considered $s$-wave scattering between
particles, assuming that the interparticle potential is a
point-like. However, as we have already pointed out,
for the potential with a finite range, the problem will always contain
scattering having an arbitrary orbital momentum $\ell$ whose
amplitude for particles having momenta equal to the
Fermi momentum $p_F$ may be estimated as
$f_l \sim a (ap_F)^{2l}$ \cite{Landau3}.
From this it follows that with the required accuracy
we can confine our analysis to $p$-wave scattering ($\ell = 1$). In
this case, only two contributions will be important: a
contribution of the order $\lambda^3$ from $p$-wave scattering at the
bare interparticle potential and a contribution of the
order $\lambda^4$ corresponding to the diagram in Fig. \ref{fig:2order}d where
one of the dashed lines corresponds to $s$-wave scattering and
the other to $p$-wave scattering. More precisely, if the amplitude
of $p$-wave scattering of two particles having momenta $p_F$
is written in the form
\begin{equation}\label{a1_scattering}
  f_1 = \alpha_1 a \left(
\frac{2 a p_F}{\pi}
  \right)^2 = \alpha_1 a \lambda^2,
\end{equation}
$\alpha_1$ is a numerical coefficient of order unity, the contribution of
triplet scattering to the irreducible vertex $\tilde{\Gamma}_{1}$ has the
form $$ \nu_F \tilde{\Gamma}_1 =
\alpha_1 \lambda^3 (1+0.008\lambda). $$
We can see that the fourth-order contribution with
respect to $\lambda$ can be neglected and consequently the
critical temperature is given by
\begin{equation}\label{tc:a1_scattering} T_{c1} = \frac{2}{\pi} e^{C} \eps_F
  \exp \left\{ - \frac{13.0}{\lambda^2[1+(4.3+\alpha_1)\lambda]}
+\frac{42.0}{[1+(4.3+\alpha_1)\lambda]^2}
\right\}.
\end{equation}

Nevertheless, we can specify a physical situation
when the contribution of $p$-wave scattering can be neglected.
This corresponds to the case when a shallow level having
the orbital momentum $\ell = 0$ (resonance scattering)
exists in the potential. In this case, the $p$-wave scattering
amplitude is estimated as $f_1 \sim r_0 (r_0
p_F)^2$,
where $r_0$ is the radius of action of the potential while the
$s$-wave scattering length is given by
$a = (1/2 m |E|)^{1/2} \gg r_0$,
where $E$ is the discrete level energy (we assume that the
condition $|E| \gg \eps_F$ is satisfied so that $ap_F \ll 1$). Then for
$\alpha_1$ in formula (\ref{tc:a1_scattering}) we obtain the estimate
$$ \alpha_1 \sim
\left( \frac{r_0}{a} \right)^3 \ll 1, $$
and if the condition
$$ \alpha_1 \ll \lambda $$
is satisfied, the contribution of the $p$-wave harmonic of the
bare interparticle interaction can be neglected compared
with the fourth order of the effective interaction
which allows only for $s$-wave scattering.

\section{Critical temperature in a magnetic field}

In this section we study the influence of an external
magnetic field on the irreducible vertex $\tilde{\Gamma}_1$ and consequently
on the critical temperature $T_{c1}$ to within terms
of the order $\lambda^3$. As was shown in \cite{Kagan89}, in the leading
approximation with respect to $\lambda$ in the model being
studied the critical $p$-wave pairing temperature may increase
appreciably if a static magnetic field is applied to the
system. This is because for conventional singlet pairing
the role of a magnetic field is always destructive due to
the paramagnetic suppression of Cooper pairing
caused by the flipping of one of the pair spins. However
for triplet $p$-wave pairing no paramagnetic effect occurs so
that the role of the magnetic field is not clear a priori.

In our approach the mechanism for variation of $T_{c1}$
in a magnetic field is based on the magnetic field dependence
of the many-particle effects which determine the
effective interaction. On the one hand, as a result of a
difference in the number of particles (and consequently
Fermi momenta) with spins directed parallel and
antiparallel to the field, the Kohn singularity increases
sharply, causing an increase in $\tilde{\Gamma}_1$. On the other hand,
$\nu_{F\downarrow} \tilde{\Gamma}_1$
value decreases with increasing magnetic
field because of a monotonic decrease in the number of
particles with spin antiparallel to the field. (We recall
that $s$-wave scattering can only occur between Fermi particles
having different spin projections.) Competition
between these two effects leads to an sharply non-monotonic
dependence of the critical temperature on
the magnetic field (more accurately, on the degree of
polarization $\alpha=(n_{\su}-n_{\sd})
/(n_{\su}+n_{\sd})$) with pronounced
increase in $T_{c1}$ for small $\alpha$, a maximum at intermediate
$\alpha$, and a decrease for $\alpha
\rightarrow 1$. (In this case of a completely
polarized Fermi gas only $p$-wave scattering between
parallel spins can only take place.)
The dependence of $\tilde{\Gamma}$ on the polarization $\alpha$ to
within second-order terms was calculated in \cite{Kagan89}:
\begin{equation}\label{Tc:k1}
  \tilde{\Gamma}^{(2)}(\delta) = -\lambda^{2}\frac{2\ln2-1}{5}
  \frac{1}{\delta^3} \left(\ds \frac{2}{1+\delta^3}
  \right)^{2/3} \left[
  1+\frac{\delta -1}{3 (2 \ln2-1)} \Psi_{\delta}
  \right],
\end{equation}
where
$$
\Psi_{\delta} = (\delta +1) \left[
10 \ln (\delta +1) - \delta^2 - 3 \right]
+ \frac{\delta-1}{2} (\delta^{3}+2 \delta^2 + 8 \delta +4)
\ln \frac{\delta+1}{\delta-1} + \frac{6}{\delta-1}
\ln\frac{
(\delta+1)}{2},
$$

$$
\delta = \frac{p_{F \ua}}{p_{F\da}} = \left(
\frac{1+\alpha}{1-\alpha}
\right)^{1/3}.
$$

\begin{figure}[t]

 \vspace{0.05\textheight}
%\centerline{\begin{picture}(0,0)%
%\centerline{ %
%\hspace{1.5in}
\setlength{\unitlength}{4144sp}
\begin{picture}(4074,2454)(439,-3313)
\put(439,-3313){\psfig{file=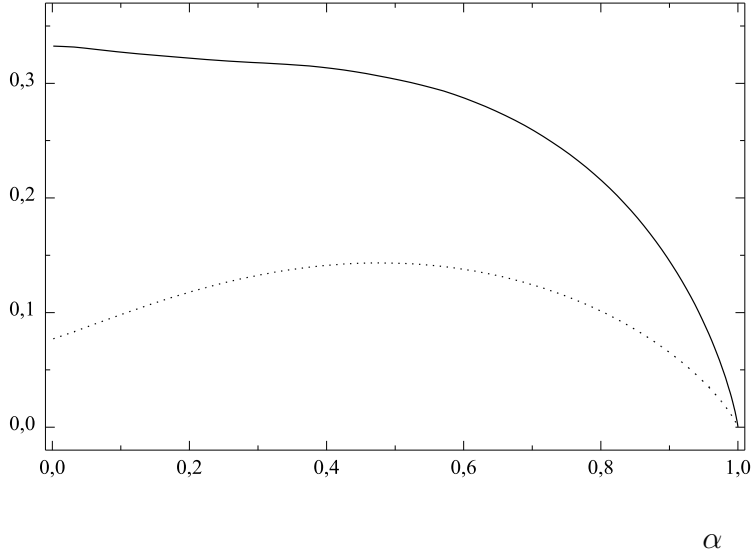,width=0.75\textwidth}}
\put(5096,-3166){\makebox(0,0)[lb]{$\alpha$}}
%\put(721,100){\makebox(0,0)[lb]{$T_{c1}/\varepsilon_F$}}
\end{picture}
%\end{picture}%
%}
\caption{
 Dependence of the second- and third-order contributions
to the irreducible vertex on the degree of polarization
$\alpha$: $\Gamma^{(2)}/\lambda^2$ -- dashed curve,
$\Gamma^{(3)}/\lambda^3$ -- solid curve.
} \vspace{0.01\textheight}
\label{fig:g2g3alpha}
\end{figure}

In the third order with respect to $\lambda$ the result can
only be obtained numerically. The corresponding contribution
is given by the diagrams in Fig. \ref{fig:3order} where the
spins on the outer lines are directed parallel to the field
and those in the inner loops can be oriented either parallel
or antiparallel to the field. The calculations
(including renormalization of the diverging diagrams)
are exactly the same as in the absence of a
magnetic field and the result is shown in Fig. \ref{fig:g2g3alpha} (solid curve)
which also contains the second-order contribution (\ref{Tc:k1}) for
comparison (dashed curve).

It can be seen that the maximum of $\tilde{\Gamma}_1^{(2)}(\alpha)$ is
obtained at $\alpha_m =
0.48$ whereas $\tilde{\Gamma}_1^{(3)}(\alpha)$ decreases
monotonically.
\begin{figure}[t]
\vspace{0.05\textheight}
%\centerline{\begin{picture}(0,0)%
%\centerline{ %
%\hspace{1.5in}
%}
%\end{picture}%
%}
\setlength{\unitlength}{4144sp}
%\centerline{
\begin{picture}(4074,2954)(439,-3313)
\put(439,-3313){\psfig{file=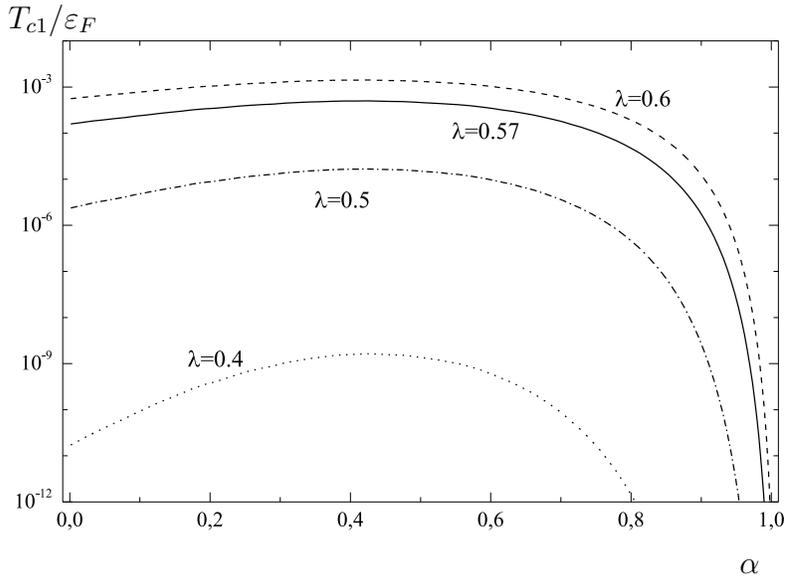,width=0.75\textwidth}}
\put(5096,-3166){\makebox(0,0)[lb]{$\alpha$}}
\put(721,100){\makebox(0,0)[lb]{$T_{c1}/\varepsilon_F$}}
\end{picture}
%}
\caption{
Dependence of $T_{c1}/ \eps_F$ on the degree of polarization
$\displaystyle \alpha $ for various $\lambda$.
}\label{fig:tclog0}
\end{figure}
Thus, the maximum of $T_{c1}$ is determined by
competition
between increasing $\tilde{\Gamma}_1^{(2)}(\alpha)$ and decreasing
$\tilde{\Gamma}_1^{(3)}(\alpha)$. For typical $\lambda$ this is in the region of $\alpha\sim 0.4$.
Graphs of the critical temperature as a function of the
degree of polarization are shown in Fig. \ref{fig:tclog0} for typical
values of $\lambda$. For $\lambda=0.6$ the value of $T_{c1}$ at the maximum
is approximately six times large than the value of
$T_{c1}$ in the absence of
the field. In the latter case the maximum is mainly determined
by the second order and is reached at $\lambda \sim 0.45$.

\section{Discussion of the results}

The experimental search for nontrivial pairing with $l \ne q 1$
in isotropic Fermi systems has recently been actively
pursued. Until recently the main candidate was \he3he4 mixture.
So far superfluidity has not yet been
observed in this system although temperatures of the order
of 97 $\mu$K have been achieved experimentally
\cite{Oh94}. In the concentration range $x<x_0\approx 3 $\% the scattering
length in the mixture correspond to the attraction so that singlet $s$-wave pairing
may be achieved. The critical temperature is
given by formula (\ref{Tc:gorkov}) allowing for
$$\varepsilon_F=\varepsilon_{F0}x^{2/3},\mbox{~}p_F=p_{F0}x^{1/3},$$
where $\varepsilon_{F0}$,\mbox{~}$p_{F0}$ are the Fermi energy and
Fermi momentum
of pure $^3$He. According to estimates made in \cite{Bashkin92}, we
have
$$\max T_{c0} = T_{c0}(1\%)\approx 10^{-4}\mbox{K}.$$
The authors of [13] predict an even lower critical temperature:
$$\max T_{c0} = T_{c0}(2\%)\approx 4\cdot 10^{-6} - 10^{-5}\mbox{K}.$$
Note that the value $T_{c0} \approx 10^{-5}$К was extracted from spin-diffusion
experiments in \cite{Bedell89} as
a fitting parameter to describe magnetostriction experiments
and $T_{c0} \approx 4 \cdot 10^{-6}$K was obtained in spin diffusion
experiments. It should be noted that for a given
concentration $x$ the gas parameter of the theory $ap_{F0}x^{1/3}$
depends weakly on pressure. Hence the pressure cannot
be considered as an instrument to obtain optimum
parameters for $s$-wave pairing.

\begin{figure}[t]
 \vspace{0.05\textheight}
%\centerline{\begin{picture}(0,0)%
%\centerline{ %
%\hspace{1.5in}
\setlength{\unitlength}{4144sp}
\begin{picture}(4074,2454)(439,-3313)
\put(439,-3313){\psfig{file=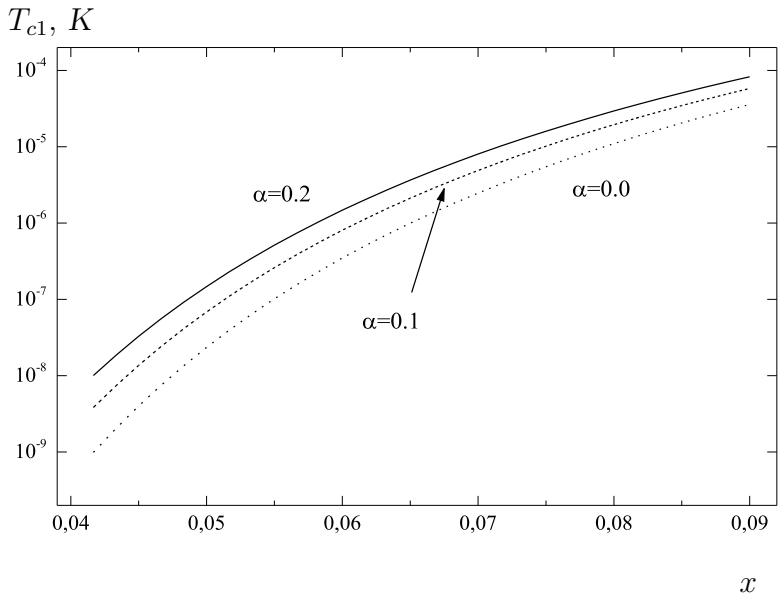,width=0.75\textwidth}}
\put(5096,-3066){\makebox(0,0)[lb]{$x$}}
\put(721,300){\makebox(0,0)[lb]{$T_{c1}$, {\sl K}}}
\end{picture}
%\end{picture}%
%}
\caption{
 Dependence of $T_{c1}$ on the concentration $x$ in a \he3he4
 mixture for various degrees of polarization:
$\alpha=0.2$ -- solid curve, $\alpha=0.1$ -- dashed curve, and $\alpha=0.0$ --
dotted curve.}
\label{fig:tcx}
\end{figure}

For high concentrations ($x>x_0$) the scattering length
changes sign $a>0$ and $s$-wave pairing becomes impossible.
Nevertheless, in this case the subsystem of $^3$He atoms
may become superfluid, but now with respect to
 $p$-wave pairing. The critical temperature is
given by formula (\ref{Tc_all:expand2}) with $\lambda$ replaced by $\lambda x^{1/3}$
 and $\eps_{F}$
replaced by $\eps_{F0}x^{2/3}$. It has a maximum at $P=10$ atm
when the maximum $^3$He concentration of 9.5\% is
achieved. Figure \ref{fig:tcx} gives the dependence of $T_{c1}$ on the
concentration calculated by using the extrapolation formula (\ref{Tc_all:expand2}).
At maximum concentration $x=9.5\%$ the
temperature $T_{c1}$ is of the order of $10^{-5}$К. A further
increase in $T_{c1}$ in solution may occur in strong magnetic
field. For example, at $x=9.5$\% the maximum of $T_{c1}$ in
a field is more than six times that in the absence of a
field, leading us to experimentally measurable temperatures
of $6 \cdot 10^{-5}$К.

Recently the properties of trapped Bose-condensed gases of
alkali elements ($^{23}$Na,
$^7$Li, $^{87}$Rb) have been
studied intensively. A combination of laser and evaporative
cooling in magnetic traps can reach gas-phase
densities of the order of $10^{12} -  10^{14}$ cm$^{-3}$ and temperatures
of the order
of $10^{-6} -  10^{-8}$ К. In addition these elements may have a
anomalously large scattering length $a$ of quasi-resonant origin. For Rb
and Na the scattering lengths are positive. It is also found
that the scattering length may cover a broad spectrum of
values from negative to positive as a result of the Feshbach
effect. This effect was observed for $^{23}$Na
\cite{Ketterle99}.

\newcommand{\li}{$^6$Li}
A logical continuation of studies of Bose condensation
in trapped gases of alkali atoms would be to obtain superfluidity in
low density fermionic system in restricted geometry. The
case of a negative scattering length makes it possible to
achieve $s$-wave pairing with a transition temperature determined
by formula (\ref{Tc:gorkov}). For \li, for example, we have
$a=-2.3 \cdot 10^{-3}\AA<0$. Thus, for $n\sim 10^{14}
\mbox{cm} ^{-3}$ the critical temperature
$T_{c0}$ is of the order of $\sim 10^{-6}\mbox{K}$. Note
that because of the Pauli principle the wave function of
an $s$-wave Cooper pair should be antisymmetric with
respect to the interchange of quantum numbers characterizing
the internal state of the atoms forming the
pair. These numbers are indices determining the multiplet
component of the hyperfine interaction for the
case of zero field (optical trap) or weak magnetic field.
 They can also correspond to the projections of the nuclear
 spin when the strong external
magnetic field of the trap destroys the hyperfine coupling.
Thus, $s$-wave pairing can only take place between
atoms of different gas components. This imposes a very
stringent constraint on the closeness of their densities.
From the experimental point of view we should have:
$|n_1-n_2|/(n_1+n_2)\leq T_{c0}/\eps_F \ll 1$
In the opposite case the Cooper pair would have a
velocity higher than the critical velocity $v_c \sim T_{c0}/p_F$,
and hence will be destroyed. From the experimental
point of view, it may therefore
 prove difficult to achieve this type of
pairing experimentally.
For $p$-wave pairing, a Cooper pair may be formed by
atoms of the same component (an analog of the $A2$
phase in superfluid $^3$He). Note that the superfluid transition
temperature in the triplet case may be increased
substantially by utilizing the presence of several components
in the trap. This increase is similar to the increase
in $T_{c1}$ in a magnetic field and is associated
with the idea of the separation
of channels: Cooper pairing is achieved
between particles of one component as a result of the
polarization of the other components. In this case, it is
possible to obtain a superfluid $p$-wave pairing temperature of
the order of $10^{-7}  -  10^{-5}$К, which is quite feasible experimentally.
By virtue of this fact this type of pairing may
be quite promising from the experimental point of view.

In conclusion the authors thank D. Rainer, H.W. Capel,
G. Frossati, R. Jochemsen, A.F. Andreev, I.A. Fomin,
I.M. Suslov, Yu. Kagan, A.S. Alexandrov, and
A.A. Golubov for many useful discussions. This work
was supported financially by the Russian Foundation for
Basic Research (projects no.
98-02-17077, 97-02-16532, and 96-15-96942) and INTAS (project no. 97-0963).

\end{document}